\documentclass[fleqn,usenatbib, letters]{mnras}

\usepackage{newtxtext,newtxmath}
\usepackage[T1]{fontenc}

\DeclareRobustCommand{\VAN}[3]{#2}
\let\VANthebibliography\thebibliography
\def\thebibliography{\DeclareRobustCommand{\VAN}[3]{##3}\VANthebibliography}

\usepackage{graphicx}	% Including figure files
\usepackage{amsmath}	% Advanced maths commands
\usepackage{multirow}
\usepackage{layouts}

\graphicspath{{./}{figures/}}

\newcommand{\angstrom}{\mbox{\normalfont\AA}}

\defcitealias{Thanathibodee2023}{T23}

\title[Low accretors and disc dispersal]{Lowest accreting protoplanetary discs consistent with X-ray photoevaporation driving their final dispersal}

\author[B. Ercolano et al.]{
Barbara Ercolano,$^{1}$\thanks{E-mail: ercolano@usm.lmu.de}
Giovanni Picogna$^{1}$
and Kristina Monsch$^{2}$
\\
$^{1}$ University Observatory, Faculty of Physics, Ludwig-Maximilians-Universit\"at M\"unchen, Scheinerstr. 1, 81679 Munich, Germany\\
$^{2}$ Harvard-Smithsonian Center for Astrophysics, 60 Garden Street, Cambridge MA 02138, USA
}

\date{Accepted XXX. Received YYY; in original form ZZZ}

\pubyear{2023}

\begin{document}
\label{firstpage}
\pagerange{\pageref{firstpage}--\pageref{lastpage}}
\maketitle

\begin{abstract}

Photoevaporation from high energy stellar radiation has been thought to drive the dispersal of protoplanetary discs. Different theoretical models have been proposed, but their predictions diverge in terms of the rate and modality at which discs lose their mass, with significant implications for the formation and evolution of planets. In this paper we use disc population synthesis models to interpret recent observations of the lowest accreting protoplanetary discs, comparing predictions from EUV-driven, FUV-driven and X-ray driven photoevaporation models. We show that the recent observational data of stars with low accretion rates (low accretors) point to X-ray photoevaporation as the preferred mechanism driving the final stages of protoplanetary disc dispersal. We also show that the distribution of accretion rates predicted by the X-ray photoevaporation model is consistent with observations, while other dispersal models tested here are clearly ruled out.

%The lowest accreting protoplanetary disks show signs of a physical limit reached in new line profiles observations. The intriguing idea that we can constraint in this way the mechanism responsible for disc dispersal is tested. We compared the EUV and X-ray internal photoevaporation models in predicting the observed population while constraining its disc life-time. We find that both models can reproduce the low-accretors. However, the X-ray photoevaporation is preferred since the low-accretors are part of the bulk distribution of the predicted disc evolution with the same median accretion rate, rather than its upper tail as for the internal EUV.
\end{abstract}


%%%%%%%%%%%%%%%%%%%%%%%%%%%%%%%%%%%%%%%%%%%%%%%%%%

%%%%%%%%%%%%%%%%% BODY OF PAPER %%%%%%%%%%%%%%%%%%

\section{Introduction}

The mechanisms shaping the final stages of protoplanetary disc evolution may have a dramatic impact on the outcome of the planet formation process \citep[e.g.][]{Ercolano2015, Ercolano2017a, Monsch2019, Jennings2018, Alexander2012, Carrera2017}. Disc winds--driven by photoevaporation from the central star--are a promising candidate to explain the phenomenology of the late stages of evolution leading to the final dispersal of the disc material \citep[e.g.][]{Ercolano2016, Ercolano2017b, Weber2020}, while magnetically driven discs may play a role in the early evolution \citep[e.g.][]{Ercolano2017, Pascucci2022, Lesur2022}. 

In the viscous accretion scenario, accretion rates in protoplanetary discs are expected to decrease with age, while mass-loss rates due to X-ray or EUV-driven photoevaporation should remain roughly constant throughout the lifetime of the disc. Over the years several photoevaporation models have been developed \citep[e.g.][for recent reviews]{Ercolano2017,Ercolano2022}, which predict different values for the mass-loss rates. Photoevaporation can only be efficient at dispersing the disc when the wind rates exceed the rate at which the disc is accreting. Consequently, in this picture, the objects with the lowest possible accretion rates should be stars transitioning from disc-bearing to disc-less and thus offer an insight into the final stages of disc dispersal. 

%Circumstellar disks are studied since more than 40 years, though the mechanism responsible for their evolution and final dispersal remains elusive \citep[see for recent reviews][]{Pascucci2022,Lesur2022}. It has been determined observationally that the disc fraction of accreting pre-main-sequence stars decreases with the age of the population \citep[see e.g.][]{Mamajek2009}, while having $\sim10\%$ of non-accreting star bearing discs \citep{Skrutskie1990}.
%The transition between accreting and non-accreting stars plays an important role in determining the final architecture of the forming planetary systems, as the speed of this process can cut out the supply for forming giant planets and halt their migration \citep[see e.g.][]{Monsch2019}.

In this context, \citet{Thanathibodee2022} recently looked for an accretion signature in disc-bearing stars previously thought to be non-accretors, using the He~I $\lambda10830 \angstrom$ line. This high excitation line allowed them to probe material in the innermost regions of protoplanetary discs, possibly detecting accretion streamers.
They found that a large fraction of this sample (at least $20-30\%$) indeed shows signs of accretion via strong red-shifted absorption consistent with free-fall velocities, preferentially at young ages and almost independently of the stellar mass.
The accretion rates were then determined independently by fitting the H$\alpha$ profiles of a sub-sample of these stars using magnetospheric accretion models \citep[][from here on \citetalias{Thanathibodee2023}]{Thanathibodee2023}.
Interestingly, although the sample size was very small (24 sources), the authors derived a minimum accretion rate of the order of $10^{-10}$ M$_\odot$ yr$^{-1}$, which is roughly one order of magnitude above the detection limit for their sample. Thus the observed value for the accretion rates of the lowest accreting objects hints at a physical mechanism shutting off accretion in the last stages of evolution. 

%They further suggested that, since this rate is similar to the mass-loss rate provided by classical models of EUV photoevaporation \citep[see e.g.][]{Alexander2007}, this mechanism of disc dispersal is a viable candidate to explain the transition between an accreting and non-accreting disc.

In this work, we test three photoevaporation scenarios, namely EUV-driven \citep{Alexander2006}, FUV-driven \citep{Komaki2021}, and X-ray driven \citep{Ercolano2021,Picogna2021} photoevaporation, for which suitable mass-loss prescriptions exist. To this aim, we produce synthetic populations of viscously evolving discs dispersed by photoevaporation to produce mass accretion distributions to compare with the recent observations of \citetalias{Thanathibodee2023}. 
%The methods are described in more detail in Section~\ref{sec:methods}, while our results are presented in Section~\ref{sec:results} and summarised together with our conclusions in Section~\ref{sec:conclusions}.

\section{Methods}\label{sec:methods}
\begin{figure*}
    \includegraphics{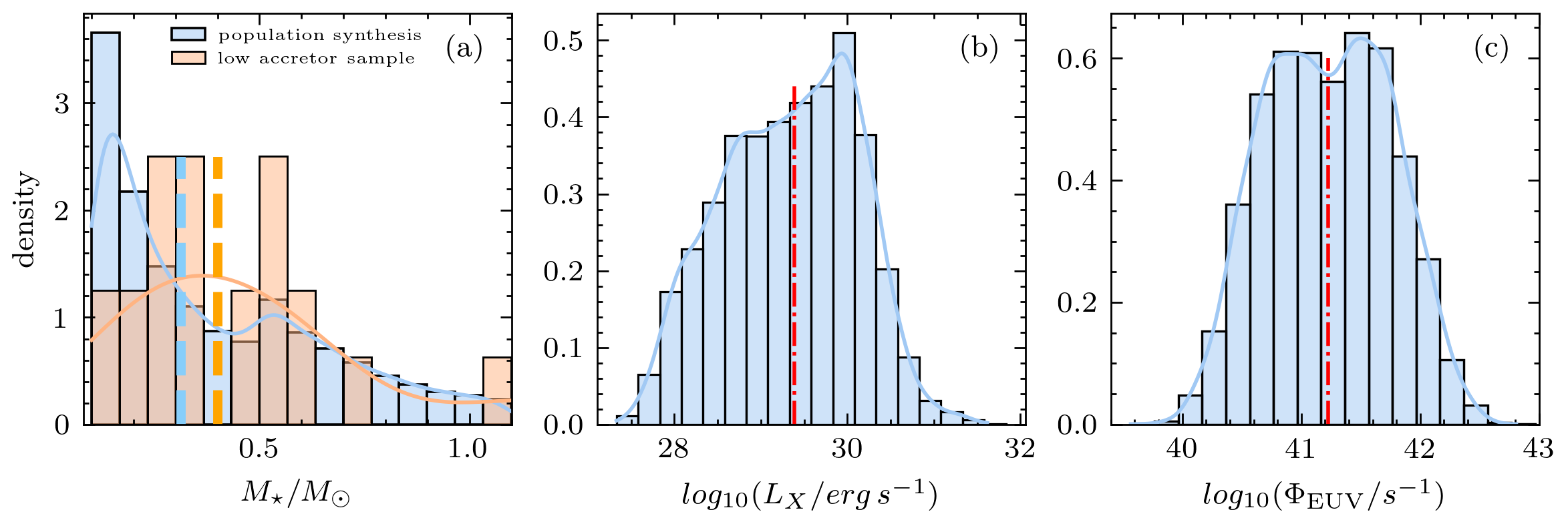}
    \caption{Panel a: histogram of the stellar mass distribution in our population synthesis (in blue) compared with the one from the sample of \citetalias[][in orange]{Thanathibodee2023}, where the KDE is overplotted with a solid blue and orange line respectively, and the median value ($0.3$ M$_\odot$) is marked with a dashed blue line for our sample and with a dashed orange line for the observed sample ($0.4$ M$_\odot$). 
    Panel b: histogram of the X-ray luminosity density distribution in the X-ray population synthesis, where the KDE is overplotted with a blue solid line and the median value of $29.4$ is marked with a dotted dashed red line. Panel c: histogram of the ionizing flux density distribution in the EUV population synthesis, where the KDE is overplotted with a blue solid line and the median value of $41.2$ is marked with a dotted dashed red line.\label{fig:hist}}
\end{figure*}

We use the one-dimensional viscous evolution code \textsc{SPOCK} \citep{Ercolano2015} which evolves the gas following
\begin{equation}
    \frac{\partial{\Sigma}}{\partial{t}} = \frac{1}{R}\frac{\partial}{\partial{R}}\left[3R^{1/2}\frac{\partial}{\partial{R}}(\nu\Sigma R^{1/2})\right] - \dot{\Sigma}_w(R,t)\,,
\end{equation}
where the first term on the right hand side describes the viscous disc evolution \citep{LyndenBell1974}, and the second one is a sink term modelling the mass-loss due to internal disc photoevaporation.

We consider either an EUV, FUV, or X-ray mass-loss rate internal photoevaporation profile as they are well-studied in the literature, and simple 1D prescriptions of the disc mass-loss rates are provided.

\subsection{X-ray photoevaporation}

The X-ray surface mass-loss rate is given following equations 2 and 3 of \citet{Picogna2019}, 
%\begin{eqnarray}
%  \label{eq:surf1}
%  \dot{\Sigma}_\mathrm{XEUV}(R) &= \ln{(10)} \bigg(\frac{6\, a\, \ln{(R)}^5}{R\, \ln{(10)}^6} +
%  \frac{5\, b\, \ln{(R)}^4}{R\, \ln{(10)}^5} +
%  \frac{4\, c\, \ln{(R)}^3}{R\, \ln{(10)}^4} + \\ \nonumber
%  &\frac{3\, d\, \ln{(R)}^2}{R\, \ln{(10)}^3} +
%  \frac{2\, e\, \ln{(R)}}{R\, \ln{(10)}^2} + \\ \nonumber
%  &\frac{f}{R\, \ln{(10)}}\bigg)
%  \frac{\dot{\mathrm{M}}_\mathrm{XEUV}(R)}{2\pi\, R} \,,
%\end{eqnarray}
%where
%\begin{equation}
%\label{eq:surf2}
%  \frac{\dot{\mathrm{M}}_\mathrm{XEUV}(R)}{\dot{\mathrm{M}}_\mathrm{XEUV}(M_\star, L_{X,\mathrm{soft}})} = 10^{a\log{R}^6 + b\log{R}^5 + c\log{R}^4 + d\log{R}^3 + e\log{R}^2 + f\log{R} + g}
%\end{equation}
with the parameters for the different stellar masses being provided in Table~1 of \citet{Picogna2021}. The mass-loss rate as a function of X-ray luminosity and stellar mass can be described by \citep{Ercolano2021,Picogna2021}
\begin{equation}
    \dot{M}_\mathrm{XEUV}(M_\star, L_{X,\mathrm{soft}}) = \dot{M}_\mathrm{XEUV}(M_\star)\frac{\dot{M}_\mathrm{XEUV}(L_{X,\mathrm{soft}})}{\dot{M}_\mathrm{XEUV}(L_{X,\mathrm{soft, mean}})} \,,
\end{equation}
where the mass-loss rate as a function of stellar mass is 
\begin{equation}
    \dot{M}_\mathrm{XEUV}(M_\star) = 3.93\times10^{-8} \left(\frac{M_\star}{M_\odot}\right)
\end{equation}
and the dependence on the soft component of the X-ray luminosity is given by
\begin{equation}
    \dot{M}_\mathrm{XEUV}(L_{X,\mathrm{soft}}) = 10^{a_L \exp{\left(\frac{(\ln(\log(L_{X,\mathrm{soft}})-b_L)^2}{c_L}\right)+d_L}} \,,
\end{equation}
with $a_L = -1.947\cdot 10^{17}$, $b_L = -1.572\cdot 10^{-4}$, $c_L = -0.2866$, $d_L = -6.694$. The soft component of the X-ray luminosity is given by
\begin{equation}
    L_{X,\mathrm{soft}} = 10^{0.95 \log{(L_X)}+1.19} \,,
\end{equation}
and was obtained by performing a linear fit between the nominal and soft X-ray luminosities listed in Table 4 of \citet{Ercolano2021}.
The mean component of the X-ray luminosity is the soft X-ray luminosity $L_{X,\mathrm{soft, mean}}$ of a star with a total X-ray luminosity given by the observational relation between stellar mass and X-ray luminosity \citep[see eq.~\ref{eq:LxMstar}, ][]{Gudel2007}.

%\begin{table*}
%\caption{Parameters for the Surface density profiles in equations~\ref{eq:surf1},%\ref{eq:surf2}}
%\label{tab:fit}
%\centering
%\begin{tabular}{c c c c c c c c c}
%\hline
%$\mathrm{M}_\star$ [$M_\odot$] & a & b & c & d & e & f & g & $\dot{\mathrm{M}}_w$ \\
%\hline
%\hline
%   $1.0$ & $-0.6344$ & $6.3587$ & $-26.1445$ & $56.4477$ & $-67.7403$ & $43.9212$ & $-13.2316$ & $3.86446$\\
%   $0.5$ & $-1.2320$ & $10.8505$ & $-38.6939$ & $71.2489$ & $-71.4279$ & $37.8707$ & $-9.3508$ & $1.9046$\\
%   $0.3$ & $-1.3206$ & $13.0475$ & $-53.6990$ & $117.6027$ & $-144.3769$ & $94.7854$ & $-26.7363$ & $1.17156$\\
%   $0.1$ & $-3.8337$ & $22.9100$ & $-55.1282$ & $67.8919$ & $-45.0138$ & $16.2977$ & $-3.5426$ & $0.37588$\\
%\hline
%\end{tabular}
%\end{table*}

\subsection{EUV photoevaporation}

We divided the EUV surface mass-loss rate in its diffuse and direct components following \citet{Alexander2007}:
\begin{equation}
    \dot{\Sigma}_\mathrm{EUV}(R) = \dot{\Sigma}_\mathrm{diffuse}(R) +  \dot{\Sigma}_\mathrm{direct}(R,t)\cdot f(R) \,,
\end{equation}
where $f(R) = 1+\exp{(-\frac{R-R_\mathrm{in}}{H_\mathrm{in}})}$ is a smoothing function to avoid numerical problems close to the disc inner edge, $R_\mathrm{in}(t)$ is the radius of the inner disc edge and $H_\mathrm{in}(t)$ is the disc scale height at the inner edge.
\begin{figure*}
    \includegraphics{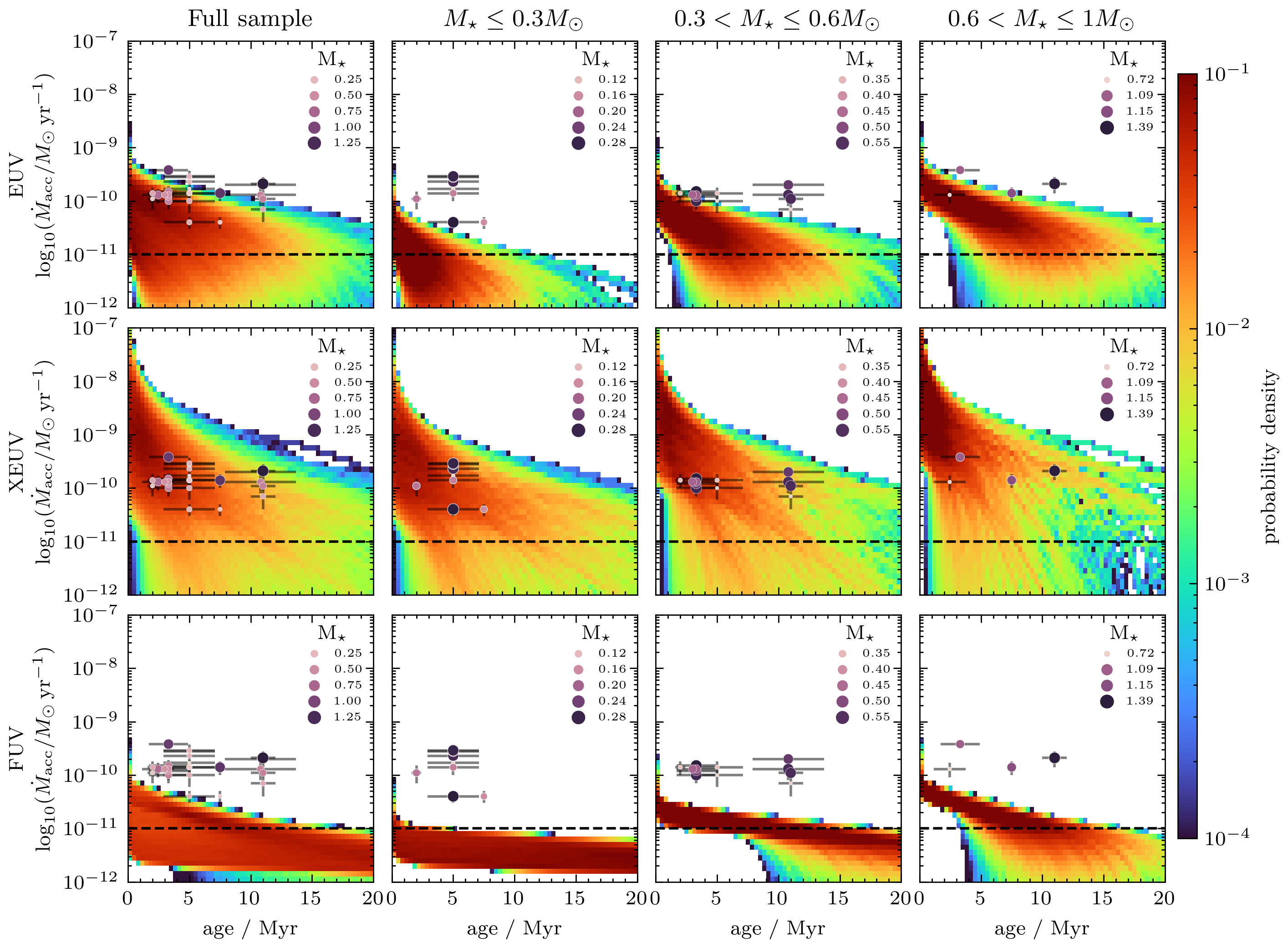}
    \caption{Synthetic populations showing the accretion rates as a function of age. The color mapping shows the probability of finding an object of a given age accreting at a given accretion rate (see text for details). Top row: discs are dispersed by EUV photoevaporation; bottom row: discs are dispersed by X-ray photoevaporation. The first column shows the full stellar mass sample, while the following columns show the result divided in three different stellar mass bins. The low-accretors population of T23 is overplotted with dots with variable size increasing by their stellar mass.
    \label{fig:mdot_age}}
\end{figure*}

The diffuse component of the EUV surface mass-loss rate is then given by
\begin{equation}
    \dot{\Sigma}_\mathrm{diffuse}(R) = 2n_0(R)u_l(R)\mu m_H \\ \mathrm{for }\ R\geq 0.1 R_g\,,
\end{equation}
where the density at the base of the flow is
\begin{equation}
    n_0(R) = C_1 \left(\frac{3\Phi_\mathrm{diff}}{4\pi\alpha_B R_g^3}\right)^{1/2} \left(\frac{2}{(R/R_g)^{15/2}+(R/R_g)^{25/2}}\right)^{1/5} \,,
\end{equation}
the wind launch velocity 
\begin{equation}
    u_l(R) = c_s A \exp{\left[B\left(\frac{R}{R_g} - 0.1\right)\right]} \left(\frac{R}{R_g} - 0.1\right)^D \,,
\end{equation}
$\mu = 1.35$ the mean molecular weight of the ionized gas, $m_H$ the mass of a hydrogen atom, $C_1 \simeq 0.14$, $R_g$ the gravitational radius, $\alpha_B = 2.6\cdot10^{-13}$ is the Case B recombination coefficient for atomic hydrogen at $10^4$ K, $c_s = 10$ km $\mathrm{s}^{-1}$ the sound speed of the ionized gas, $A=0.3423$, $B=0.3612$, $D=0.2457$, and the stellar diffuse ionizing EUV flux
\begin{equation}
    \Phi_\mathrm{diff} = 
    \begin{cases}
        \Phi_\mathrm{EUV}\left(\frac{R_\mathrm{thin}}{R_\mathrm{in}}\right) & \text{if } R_\mathrm{thin} < R_\mathrm{crit},\\
        \Phi_\mathrm{EUV} & \text{otherwise},
    \end{cases}
\end{equation}
where $\Phi_\mathrm{EUV}$ is the unattenuated stellar ionizing EUV flux (in photons s$^{-1}$), $R_\mathrm{thin}$ is the radius at which the disc becomes optically thin in the vertical direction
\begin{equation}
    \Sigma_g(R_\mathrm{thin}) = m_H \sigma_\mathrm{13.6\, eV}^{-1} \,,
\end{equation}
$\sigma_\mathrm{13.6\, eV} = 6.3\cdot10^{-18}$ cm$^2$ is the absorption cross-section for ionizing photons, $R_\mathrm{crit}$ is the critical radius at which the gap opens
\begin{equation}
    R_\mathrm{crit} = 1.4 \left(\frac{M_\star}{M_\odot}\right) \ \text{au}\,.
\end{equation}

The direct component of the EUV surface mass-loss rate (defined only for $R>R_\mathrm{in}$) is given by
\begin{equation}
    \dot{\Sigma}_\mathrm{direct}(R, t) = 2 C_2 \mu m_H c_s \left[\frac{\Phi_\mathrm{EUV}}{4\pi\alpha_B(H/R) R_\mathrm{in}^3(t)}\right]^{1/2} \left[\frac{R}{R_\mathrm{in}(t)}\right]^{-a} \,,
\end{equation}
where $C_2 = 0.235$, $a = 2.42$, $H/R$ is the disc aspect ratio.

\subsection{FUV photoevaporation}

The combined X-ray/FUV component of the surface mass-loss rate has been defined in eq. A1 of \cite{Komaki2021} with coefficients from their Table 2 (as a function of stellar mass) and their Table 3 (as a function of X-ray/FUV luminosities). However, since in their prescription the FUV contribution is significantly more important than the X-ray one, we assume here a fixed X-ray luminosity and consider only the variation of the FUV component. To simplify the implementation we considered here only the coefficients for a 1 Solar mass star with varying FUV luminosities, and scale then the total mass-loss rate as a function of stellar mass by \citep{Komaki2021}
\begin{equation}
    \dot{M}_\mathrm{FUV} \propto \left(\frac{M_\star}{M_\odot}\right)^{2.06} \,.
\end{equation}

\subsection{Population synthesis}
We assume an initial stellar mass function following \citet{Kroupa2001}, 
%\begin{equation} \label{eq:IMF}
    $\xi(m) \propto m^{-\alpha}$ ,
%\end{equation}
where $\alpha = 1.3 \pm 0.5$ for $0.08 \leq m/M_\odot < 0.5$ and $\alpha = 2.3 \pm 0.3$ for $0.5 \leq m/M_\odot \leq 1$, from which we obtain the distribution shown in Figure~\ref{fig:hist} (a) for a sample of 10,000 stars with a median value of $\sim 0.3$ M$_\odot$.
The population of observed low accretors from \citetalias{Thanathibodee2023} is overplotted in orange, where one can see that the distribution is significantly different from the adopted one due to the low number statistics (24 sources), but the mean of the distribution ($0.4$ M$_\odot$, shown with a black dashed line) is consistent with the adopted sample.

\begin{figure*}
    \includegraphics{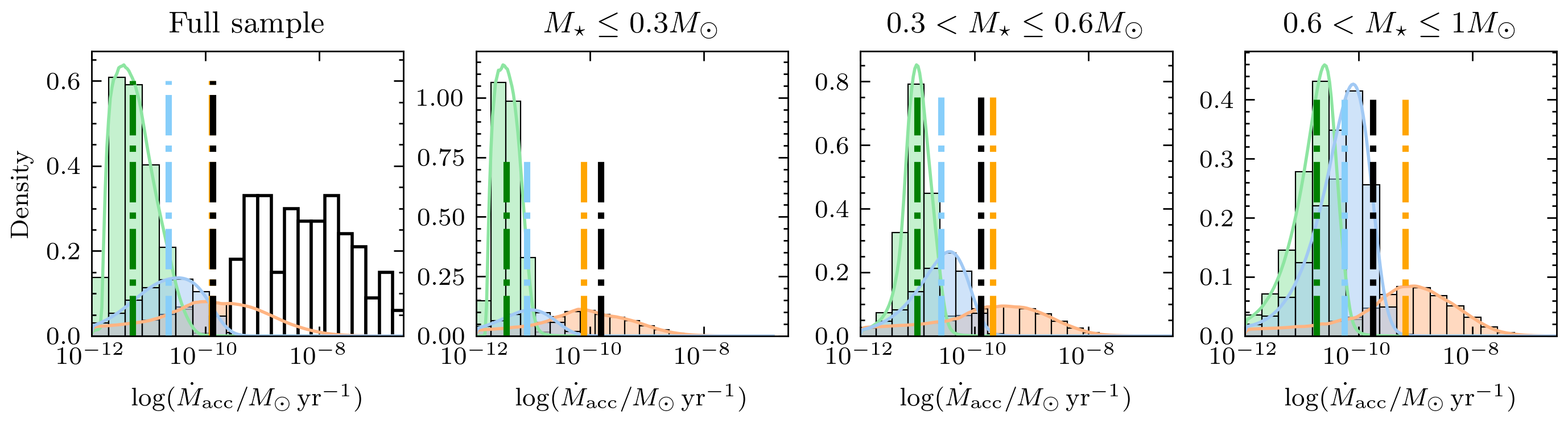}
    \caption{Histogram of the accretion rate density distribution in the EUV (blue), X-ray (orange), and FUV (green) population synthesis. The KDE is overplotted with a solid line for each distribution, and the median values are plotted with dotted-dashed lines. For direct comparison the median accretion rate of the low-accretors population is plotted with a black dotted-dashed line. The distribution of accretion rates from the sample of \citet{Manara2023} is added with a black histogram in the full sample for comparison.}\label{fig:hist_mdot}
\end{figure*}

\citet{Gudel2007} derived an observational relation between the median X-ray luminosities and stellar masses
\begin{equation} \label{eq:LxMstar}
    \log_{10}(L_{X}) = (1.54\pm0.12) \log_{10}(M_\star) + (30.31\mp0.06) \,,
\end{equation}
though a large spread is observed around the mean values, which becomes larger for small mass stars \citep[e.g.][]{Getman2022}.
\citet{Kuhn2019} took a subsample of the \textit{Chandra} Orion Ultradeep Project \citep[COUP, cf.][]{Getman2005} and stratified it in three stellar mass bins using the \citet{Baraffe1998} evolutionary models. From this sample one can derive an X-ray luminosity function (XLF) as a function of stellar mass.
%, as shown in Figure~\ref{fig:XLF}, from which one can see the increase in the median X-ray luminosity as a function of stellar mass as well as the increase spreading for lower stellar masses.
We then calculated the median stellar mass for the three stellar mass bins 
%in Figure~\ref{fig:XLF} 
given the adopted IMF, and shifted the XLF distribution to match the given value of the stellar mass. We then sampled the X-ray luminosity given the probability density corrected for the stellar mass and obtained the X-ray luminosity distribution shown in Figure~\ref{fig:hist}(b) from the 10,000 sampled stellar masses, with a median value of $10^{29.4}$ erg s$^{-1}$ and a spread over four orders of magnitude.
%\begin{figure}
%    \includegraphics[width=\columnwidth]{Fig2}
%    \caption{X-ray luminosity function for a representative sample of generic young stellar objects sample of pre-main-sequence stars detected in the Chandra Orion Ultradeep Project (\textsc{COUP}; \citet{Feigelson2005,Getman2005}), stratified in 3 stellar mass bins as in \citet{Kuhn2019}. \label{fig:XLF}}
%\end{figure}
%\begin{figure}
%    \includegraphics[width=\columnwidth]{Fig3}
%    \caption{Histogram of the X-ray luminosity density distribution in the XEUV population synthesis, where the KDE is overplotted with a dark blue solid line and the median value of 29.4 is marked with a dotted dashed black line. \label{fig:Lx}}
%\end{figure}

The EUV rates are shown to scale with the ratio of incoming ionising flux. Assuming a chromospheric origins of the EUV flux, we can then adopt the same scaling relation as eq.~\ref{eq:LxMstar}
\begin{equation} \label{eq:PhiEUVMstar}
    \log_{10}(\Phi_{EUV}) = 1.54 \log_{10}(M_\star) + 42 \,
\end{equation}
and consider a small dispersion around the mean value for each stellar mass of 0.25 dex, as shown in Figure~\ref{fig:hist} (c), which gives ionizing fluxes ranging from $10^{40}\,\mathrm{s^{-1}}$ to $10^{42.5}\,\mathrm{s^{-1}}$ with a median of $10^{41.2}\,\mathrm{s^{-1}}$.

Contributions to the FUV luminosity ($L_\mathrm{FUV}$) are believed to come both from the activity in the chromosphere ($L_\mathrm{FUV,chr}$) and accretion hotspots on the stellar surface \citep[see e.g.][$L_\mathrm{FUV,acc}$]{Gorti2009}.
The accretion component, that is more important in the early stages of evolution, can be expressed as
\begin{equation}\label{eq:fuvacc}
L_\mathrm{FUV,acc} = 10^{-2} \left(\frac{M_\star}{M_\odot}\right)\left(\frac{R_\star}{R_\odot}\right)^{-1}\left(\frac{\dot{M}_\mathrm{acc}}{10^{-8} \ \mathrm{M}_\odot \mathrm{yr}^{-1}}\right) \ \mathrm{L}_\odot\,,
\end{equation}
assuming the accretion luminosity as a black-body with $T=9000$ K which has a fraction of $4\%$ in the FUV band \citep{Gorti2009}.
The chromospheric component of the FUV flux is given by
\begin{equation}\label{eq:fuvchr}
L_\mathrm{FUV,chr} = 10^{-3.3} L_\star\,,
\end{equation}
obtained for non-accreting, weak-line T Tauri stars \citep{Valenti2003}.

%shown in Figure~\ref{fig:param_euv}.
%\begin{figure*}
%    \includegraphics[width=0.92\textwidth]{Fig5}
%    \caption{KDE plot of the alpha - disc mass (left) and alpha-r1 (right) parameters that were adopted in the EUV population synthesis in order to have a disc life-time compatible with the observationally derived one. \label{fig:param_euv}}
%\end{figure*}
We constrained the disc properties in order to match the observed mean disc lifetime of 2-3 Myr \citep[see e.g.][]{Ribas2014}. For the EUV profile we sampled the viscous $\alpha$-parameter and scaling radius $R_1$ for disc masses ranging from $0.01$--$0.1 M_\star$ for a typical star with median values from our distribution ($M_\star = 0.3 M_\odot$, $\Phi_\mathrm{EUV} = 10^{41.2}$ s$^{-1}$) and then we selected those returning a disc life-time in agreement with observations and obtained a best linear fit given by:
\begin{equation}\label{eq:r1_alpha}
    R_1 = 60.4 \log_{10}({\alpha}) + 3009.7 M_d [M_\star] + 226.6 \\\text{[au]} \,.
\end{equation}
We then sampled the whole stellar mass range and obtained the full sampling of the parameter space.
For the FUV profile, we did the same procedure, finding a new best fit:
\begin{equation}\label{eq:r1_alpha_Md}
    \alpha = 10^{-9.41+1.678\log_{10}(R_1)-1.211\log_{10}(M_d [M_\star])}\,.
\end{equation}
For the X-ray profile, the disc evolution is primarily driven by the internal photoevaporation rather than the disc properties, thus we sample uniformly the parameter space fixing only the disc mass to $0.1 M_\star$, which is reasonable for non self-gravitating discs. We summarize the probed parameter space in Table~\ref{tab:popsynthtable}.
\begin{table}
	\centering
	\caption{Parameters for the population synthesis calculations.}
	\label{tab:popsynthtable}
	\begin{tabular}{lccccr}
        \hline
		\hline
		  Name & Viscosity & $R_1$ & Disc mass & Stellar mass & Stellar flux \\
        & $\log_{10}(\alpha)$ & au & $M_\star$ & $M_\odot$ & $(\Phi,\ L_X)$\\
		\hline
		  EUV & eq.~\ref{eq:r1_alpha} & eq.~\ref{eq:r1_alpha} & eq.~\ref{eq:r1_alpha} & Fig.~\ref{fig:hist} (a) & Fig.~\ref{fig:hist} (c)\\
		X & [-4, -2] & [10, 100] & 0.1 & Fig.~\ref{fig:hist} (a) & Fig.~\ref{fig:hist} (b)\\
        FUV & eq.~\ref{eq:r1_alpha_Md} & eq.~\ref{eq:r1_alpha_Md} & eq.~\ref{eq:r1_alpha_Md} & Fig.~\ref{fig:hist} (a) & variable\\
		%3 & 5 & 7 & 9\\
		\hline
	\end{tabular}
\end{table}

\section{Results}\label{sec:results}

Figure~\ref{fig:mdot_age} shows the accretion rates as a function of time for a population synthesis of 10,000 discs sampled as described in Section~\ref{sec:methods}.
Overplotted with dots of variable size (based on their stellar mass) is the population of low accretors from \citetalias{Thanathibodee2023}, and the observational limit of the He I $\lambda10830$ marked with a black dashed line at $10^{-11}$\,M$_\odot$\,yr$^{-1}$.
From the full sample, shown in the first column, one can immediately see that while the population of discs with the X-ray profile (second row) catches all the observed data points in the region with high density ($> 10^{-2}$), the populations using the EUV and FUV photoevaporation profiles (first and last row) cannot explain the data for the older star-forming regions (Orion OB1a and Upper Sco). Furthermore the observational data points lie in the upper region of the distribution, even though they should be a sample of the population accreting at the lowest possible rates. This means that the bulk of the discs in the studied star forming regions covers a region not explained by the EUV or FUV photoevaporation profiles.

Low-mass stars (small dots) cover the lower part of the high density distribution while higher-mass stars the top part. This is expected from the observationally derived relation between accretion rate and stellar masses, that shows a sharp increase of the accretion rate as a function of stellar mass ($\dot{M}$-$M_*$) with a broken power law \citep[e.g.][]{Alcala2017}. \citet{Ercolano2014} proposed that the $\dot{M}$-$M_*$ distribution is also a consequence of discs being dispersed by X-ray photoevaporation, which fits well in this overall picture. 

In Figure~\ref{fig:hist_mdot} we plot the histogram of the accretion rate distribution for the EUV (in blue) and X-ray (in orange) populations. Overplotted are the median accretion rates in dotted-dashed blue and orange lines respectively, and the median accretion rate for the observed population of low-accretors with a dotted-dashed black line. From this one can directly visualise how the median of the X-ray photoevaporating population and the observed low-accretors is very similar. On the other hand, for the population dispersing via EUV-photoevaporation the median accretion rate of the low-accretors population is close to the high-accretor wing of the EUV population, which is in tension with the observation that the observed objects should represent the lowest end of the mass accretion rate distribution.

Overplotted in Figure~\ref{fig:hist_mdot} (black hystogram) is the observed distribution of accretion rates from the sample of \citet{Manara2023} recently reported by \citet{Alexander2023}. Our results show that the observations favour X-ray photoevaporation as the main dispersal mechanism over other models, as it represents well the low-accretor population.

We finally compared the distribution of disc fractions as a function of age, calculated as the time at which the accretion rate drops below the observational limit quoted by \citetalias{Thanathibodee2023} ($10^{-11}$ M$_\odot$ yr$^{-1}$) in our population synthesis with the observed distribution of disc fraction (Figure~\ref{fig:frac_time}, black lines). The EUV and X-ray distributions fit the observational data points perfectly by construction, as we choose the parameter space in order to fit them for the median stellar mass and EUV ionising field. The FUV distribution fits less well the observed population as the stellar luminosity (and thus the wind mass-loss rate) evolves with time (see eq.~\ref{eq:fuvacc}) inducing a stronger stellar mass dependance. The main difference between the models is the parameter space adopted to match the observations which is much more limited for the EUV and FUV profiles with respect to the X-ray one (see Table.~\ref{tab:popsynthtable}).

\subsection{Mass dependent results}
\begin{figure}
    \includegraphics{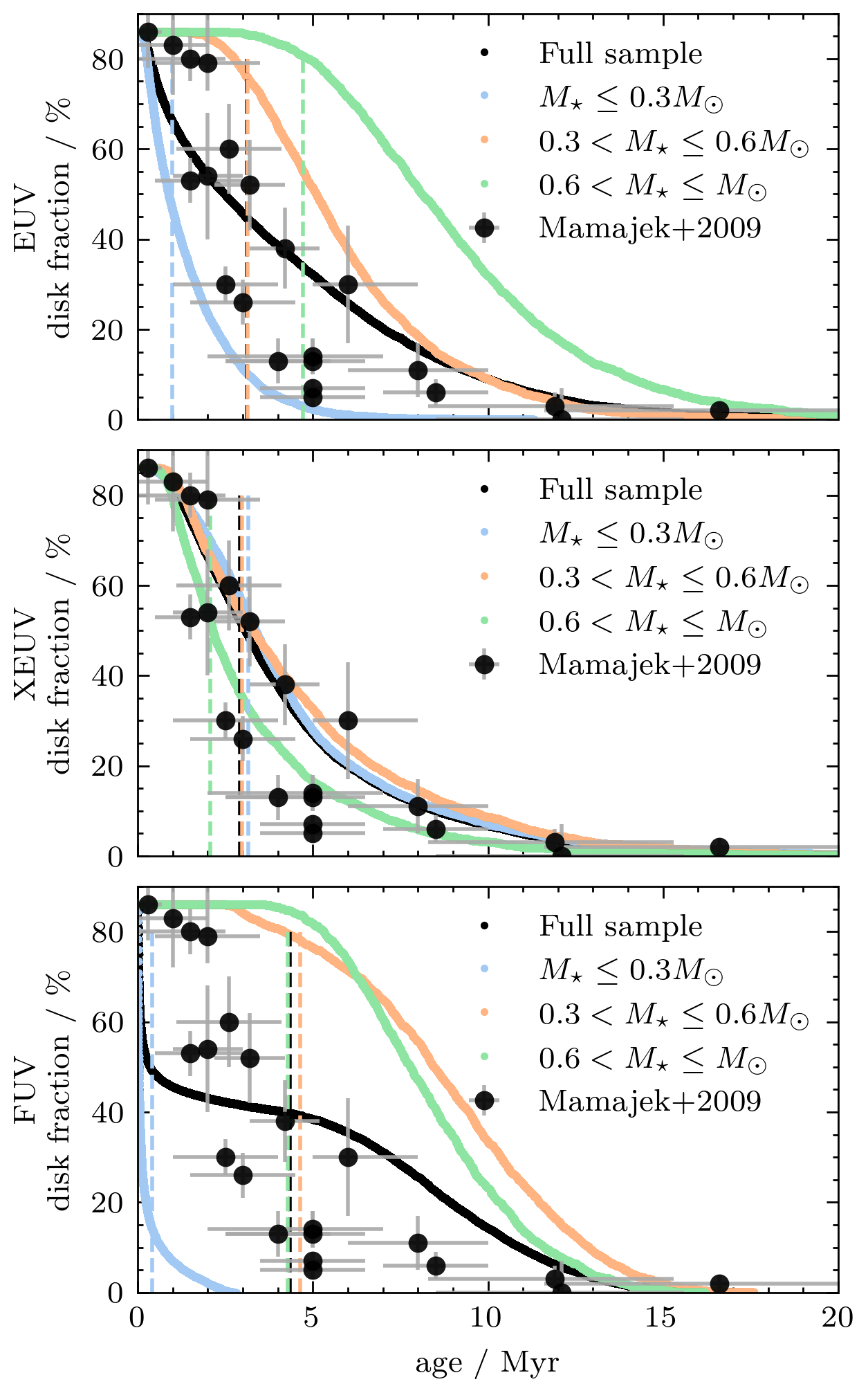}
    \caption{Disc fraction as a function of disc lifetime for the EUV (top panel), X-ray (middle panel) and FUV (bottom panel) synthetic populations compared with the observational data from \citet{Mamajek2009}, divided in three stellar mass bins. The median disc life-time increases with stellar mass for the EUV and FUV models, while it decreases for the X-ray models. \label{fig:frac_time}}
\end{figure}

The results shown in the previous section assume a well-sampled IMF, which implies that objects at the lower end of the mass range are the most abundant. However the sample of low accretors from \citetalias{Thanathibodee2023} includes objects with masses ranging between $0.1$ and $1.39$ M$_\odot$ and a mean of $0.4$ M$_\odot$, resulting in a slightly different stellar mass function with respect to the adopted IMF (see Figure~\ref{fig:hist}, a). Despite the low number of objects in observational sample it is worth trying to decouple the effect of stellar mass from the analysis. To this aim we repeat the analysis for separate mass bins, as shown in columns 2 to 4 of Figures~\ref{fig:mdot_age} and~\ref{fig:hist_mdot}. The observed median of the observed accretion rate distribution is always close the median or the lower end of distribution predicted by the X-ray model. For the EUV and FUV distributions, in contrast, the median is always close to the higher end of the predicted distribution, for all mass bins, implying that the EUV and FUV models predict that objects with accretion rates one or two order of magnitude lower than observed should be more common.

%\km{You describe this plot before. In general I find it a bit confusing that you only described one panel of each figure before going into the mass dependency. I would probably mention this approach to decouple Mstar from everything else in the very beginning, and then describe each plot fully rather than going back and forth. I think many of my comments above actually already apply to the mass-dependent results because I didn't understand that we're not talking about them yet :D}

Finally, in Figure~\ref{fig:frac_time} we explore the effect of stellar mass on the disc lifetimes. In the top panel the population with the X-ray profile profile shows little variation as a function of stellar mass, with a trend of decreasing lifetimes for increasing stellar masses, which is consistent with current observations \citep{Ribas2015}. In the bottom panel, the population with the EUV profile shows a much stronger dependence on the stellar mass, with an inverse trend of decreasing disc lifetime for smaller mass stars. The overall distribution in this case fits the observational data because of the adopted IMF, but it would overpredict the disc fractions for the stellar mass distribution studied in \citetalias{Thanathibodee2023}.

\section{Conclusions}\label{sec:conclusions}

We conducted a population synthesis study of viscously evolving discs subjected to different photoevaporation prescriptions, in order to test these models against the recently published observations of the lowest accretion rates measured by \citetalias{Thanathibodee2023}. For the case of X-ray photoevaporation the probability of finding an object at a given accretion rate changes as a function of age of the population. However for objects older than approximately 1~Myr it is clearly not improbable to find objects at rates of order $10^{-10}M_{\odot}/yr$. In particular for objects older than about 5 Myr the probability distribution peaks indeed at these values. For lower accretion rates the probabilty decreases sharply for all ages, indicating that the lowest accretors of a population of discs being dissipated by X-ray photoevaporation is indeed expected to be observed accreting at rates of order $10^{-10} M_{\odot}/yr$. 

Our findings can be synthesised as follows:
\begin{itemize}
  \item EUV, FUV and X-ray photoevaporation predict objects with accretion rates as those observed in the low-accretor sample of T23.	
	\item However, the low-accretors are part of the bulk or lower end of the distribution predicted by X-ray photoevaporation, while they fall in the upper end of the EUV and FUV driven populations, which thus would predict an overabundance of objects accreting at significantly lower rates, which is not observed. 
 \item The constraints required by EUV and FUV internal photoevaporation to explain the observed disk lifetime are considerably more stringent than those of X-ray-driven disc evolution (see eq.~\ref{eq:r1_alpha}).
	\item The relationship between disk fraction and stellar mass exhibits a strong dependence for EUV and in particular for FUV profiles, in contradiction with the observed low-accretor population that shows similar accretion rates across a wide range of stellar masses.
 \item The observed distribution of accretion rates favours X-ray photoevaporation as the main dispersal mechanism. 
 
 \end{itemize}
 
In light of the results of the population synthesis models shown here, we conclude that the recent low-accretor observations point to X-ray photoevaporation as driving the final stages of disc dispersal.

\section*{Acknowledgements}
We thank the anonymous referee for a constructive report. 
We acknowledge support of the DFG (German Research Foundation), Research Unit ``Transition discs'' - 325594231 and of the Excellence Cluster ORIGINS - EXC-2094 - 390783311. KM was supported by NASA {\it Chandra} grants GO8-19015X, TM9-20001X, GO7-18017X, and HST-GO-15326.
%%%%%%%%%%%%%%%%%%%%%%%%%%%%%%%%%%%%%%%%%%%%%%%%%%
\section*{Data Availability}

The data for the population synthesis calculations is available at \href{https://github.com/GiovanniPicogna/low-accretors}{https://github.com/GiovanniPicogna/low-accretors}


\bibliographystyle{mnras}
\bibliography{lib}

%%%%%%%%%%%%%%%%%%%%%%%%%%%%%%%%%%%%%%%%%%%%%%%%%%

%%%%%%%%%%%%%%%%% APPENDICES %%%%%%%%%%%%%%%%%%%%%

\appendix

%%%%%%%%%%%%%%%%%%%%%%%%%%%%%%%%%%%%%%%%%%%%%%%%%%

% Don't change these lines
\bsp	% typesetting comment
\label{lastpage}
\end{document}